\documentstyle[sprocl,epsf]{article}

\newdimen\rotdimen
\def\vspec#1{\special{ps:#1}}
\def\rotstart#1{\vspec{gsave currentpoint currentpoint translate
   #1 neg exch neg exch translate}}
\def\rotfinish{\vspec{currentpoint grestore moveto}}
\def\rotr#1{\rotdimen=\ht#1\advance\rotdimen by\dp#1%
   \hbox to\rotdimen{\hskip\ht#1\vbox to\wd#1{\rotstart{90 rotate}%
   \box#1\vss}\hss}\rotfinish}
\def\rotl#1{\rotdimen=\ht#1\advance\rotdimen by\dp#1%
   \hbox to\rotdimen{\vbox to\wd#1{\vskip\wd#1\rotstart{270 rotate}%
   \box#1\vss}\hss}\rotfinish}%
\def\rotu#1{\rotdimen=\ht#1\advance\rotdimen by\dp#1%
   \hbox to\wd#1{\hskip\wd#1\vbox to\rotdimen{\vskip\rotdimen
   \rotstart{-1 dup scale}\box#1\vss}\hss}\rotfinish}%
\def\rotf#1{\hbox to\wd#1{\hskip\wd#1\rotstart{-1 1 scale}%
   \box#1\hss}\rotfinish}%

\catcode`@=11
\def\citenum#1{{\def\@cite##1##2{##1}\cite{#1}}}
\catcode`@=12

\def\E{\,\rlap/\!E_T} 
\newcommand{\alt}{\mathrel{\raisebox{-.6ex}{$\stackrel{\textstyle<}{\sim}$}}}
\newcommand{\agt}{\mathrel{\raisebox{-.6ex}{$\stackrel{\textstyle>}{\sim}$}}}

\begin{document}

\vspace*{-.5in}

\font\fortssbx=cmssbx10 scaled \magstep1
\hbox to \hsize{
\hbox{\fortssbx University of Wisconsin - Madison}
\hfill$\vtop{\normalsize\hbox{\bf MADPH-98-1069}
                \hbox{July 1998}}$ }

\vspace{.2in}

\title{\uppercase{Higgs Physics and Supersymmetry}\footnote{Talk presented at the {\it Richard Arnowitt Fest: A Symposium on Supersymmetry and Gravitation}, College Station, Texas, April 1998.}}
\author{\unskip\smallskip V. BARGER}
\address{Physics Department, University of Wisconsin, Madison, WI 53706}

\maketitle
\thispagestyle{empty}
\abstracts{
The quest for the physics underlying the breaking of the electroweak symmetry and the generation of mass is surveyed.
}

\section{Introduction}

It is a pleasure to contribute to this tribute to Richard Arnowitt, whose influence over the course of particle physics has been vast. My review of the present focus of the field relies to a great extent on the supergravity theory which he and his collaborators originated.\cite{arno,hall,barbi}

The Standard Model (SM) of particle physics provides a unified understanding of the strong, weak, and electromagnetic forces. There is only one parameter, the gauge coupling, for each of the 3 simple gauge groups,\break SU(3)$\times$SU(2)$\times$U(1).  Universal interactions arise from the covariant derivative $D^\mu = \partial^\mu + ig_iA_i^\mu$ in the kinetic term of the Lagrangian. For the theory to be renormalizable, all particles start off massless, and new forces associated with a Higgs field are introduced to break the electroweak symmetry via a $\mu^2(\Phi^\dagger\Phi) + \lambda(\Phi^\dagger\Phi)^2$ interaction and to give fermion masses via Yukawa interaction $\lambda_f \bar\Psi_R \Phi^\dagger   \Psi_L + \rm h.c$. The particle states of the SM with massive neutrinos are summarized in Table~1 following the format of Ref.~\citenum{hikasa}; all have been found but the Higgs boson, which is a vital component of the theory.

The SM is a pillar of success as an effective theory. Table~2 summarizes the status of measurements reported at the 1998 Moriond conference.\cite{moriond} Here ``pull'' denotes the standard deviations from a global fit of all data to the SM. There is glorious overall agreement.

\begin{table}[t]
\caption{The particle states of the Standard Model.}
\smallskip
\centering\leavevmode
\epsfxsize=2.25in\epsffile{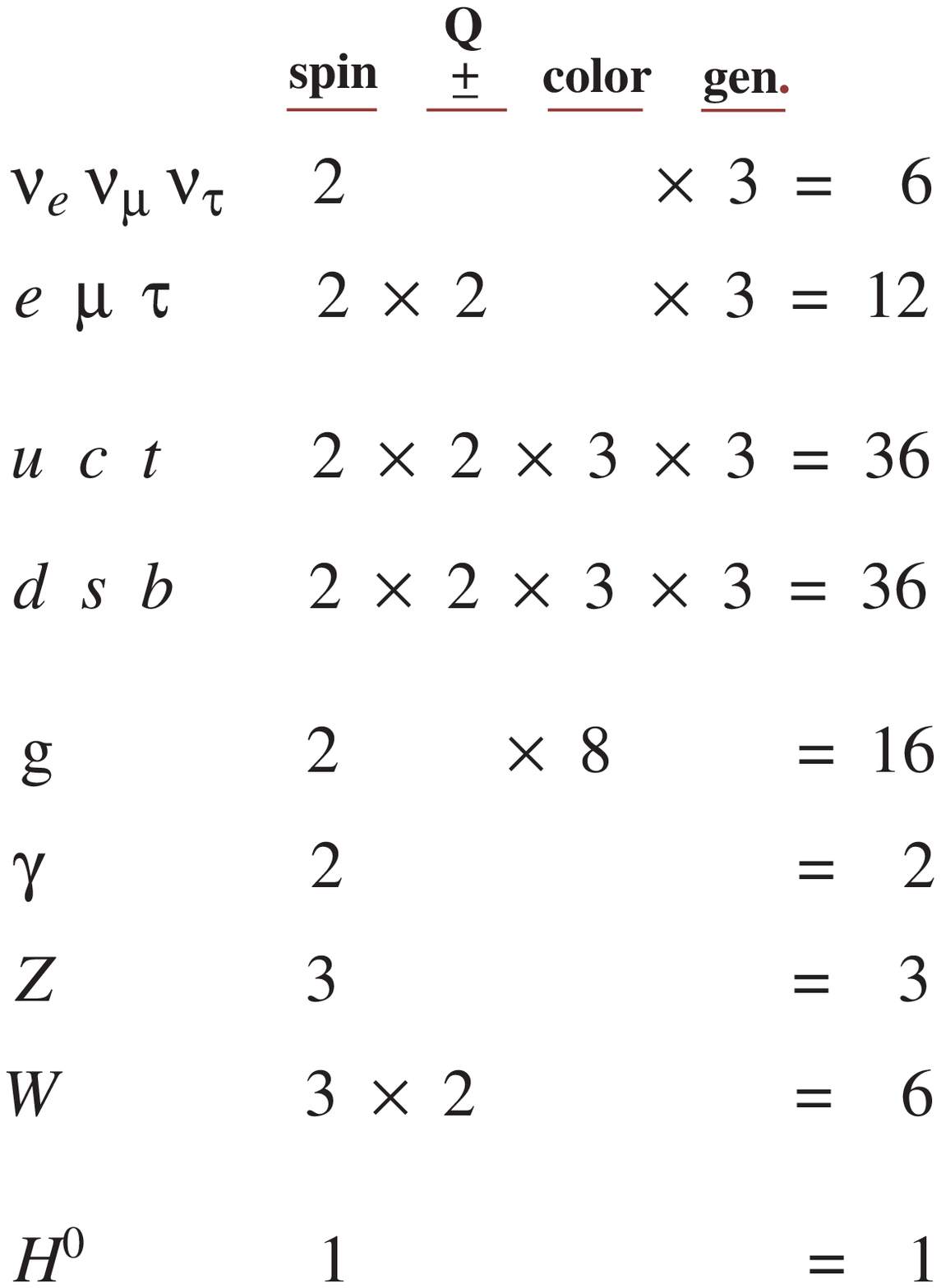}
\end{table}

\begin{table}[t]
\caption[]{Global fit to precision electroweak data. From Ref.~\citenum{moriond}.}
\smallskip
\centering\leavevmode
\epsfxsize=2.75in\epsffile{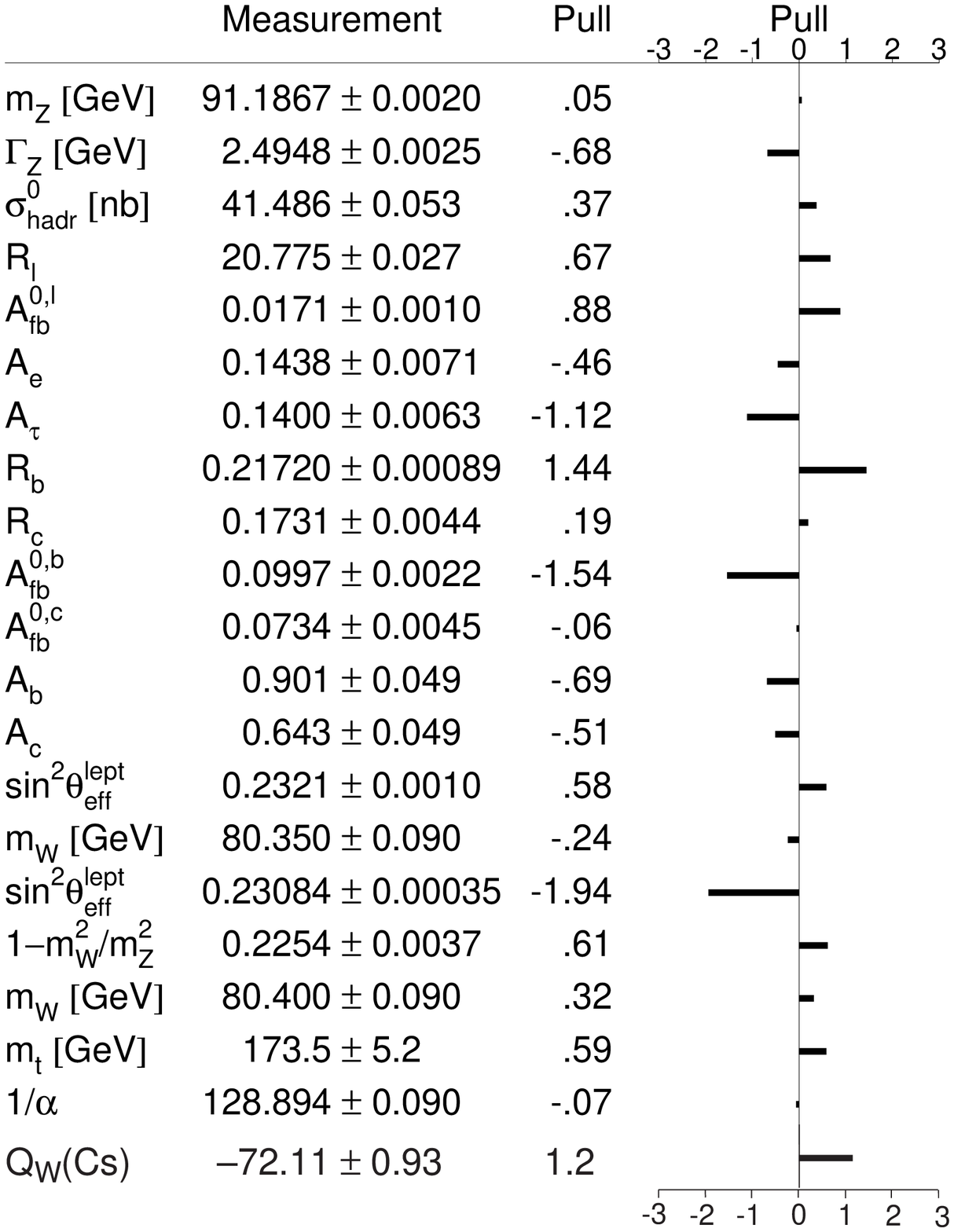}
\end{table}

\section{SM Higgs Boson}

The Higgs boson predicted by the theory is a neutral elementary particle with spin zero. Although its mass is unknown, its interactions with other particles are precisely predicted; the $H$ couplings are proportional to $m_f$ (fermion mass) and $M_W^2$ (gauge boson mass-squared).

Direct searches for the SM Higgs at the LEP-2 $e^+e^-$ collider have placed a lower limit $m_H > 89.3$~GeV on its mass. The production mechanism and search topologies are illustrated in Fig.~1.

Indirect evidence for the Higgs boson is obtained from global fits\cite{moriond,lep2-smh,mh-ind} to electroweak observables, which depend on $\log m_h$ through radiative corrections. From a fit to all data (LEP, SLD, $M_W$, $\nu N$, $m_t$), a SM Higgs mass value $m_H = 47^{+84}_{-45}$~GeV is inferred\cite{moriond,lep2-smh} ($m_H<330$~GeV at $1.96\sigma$). This result is very encouraging.

At the upgraded Tevatron, the Large Hadron Collider (LHC), and other future colliders there are a number of ways that the Higgs boson may be produced, as illustrated in Fig.~2.  The approximate ranges of the Higgs mass coverage\cite{haber} are drawn in Fig.~3. The bottom line is that the SM Higgs will not escape detection. The near term focus is Higgs searches at LEP-2 with increased c.m.\ energy $\sqrt s$. Figure~4 shows the expected Higgs discovery potential\cite{bighiggs} at $\sqrt s = 192$ and 205~GeV (LEP-2 is currently running at 189~GeV). 

\begin{figure}[t]
\centering\leavevmode
\epsfxsize=2.75in\epsffile{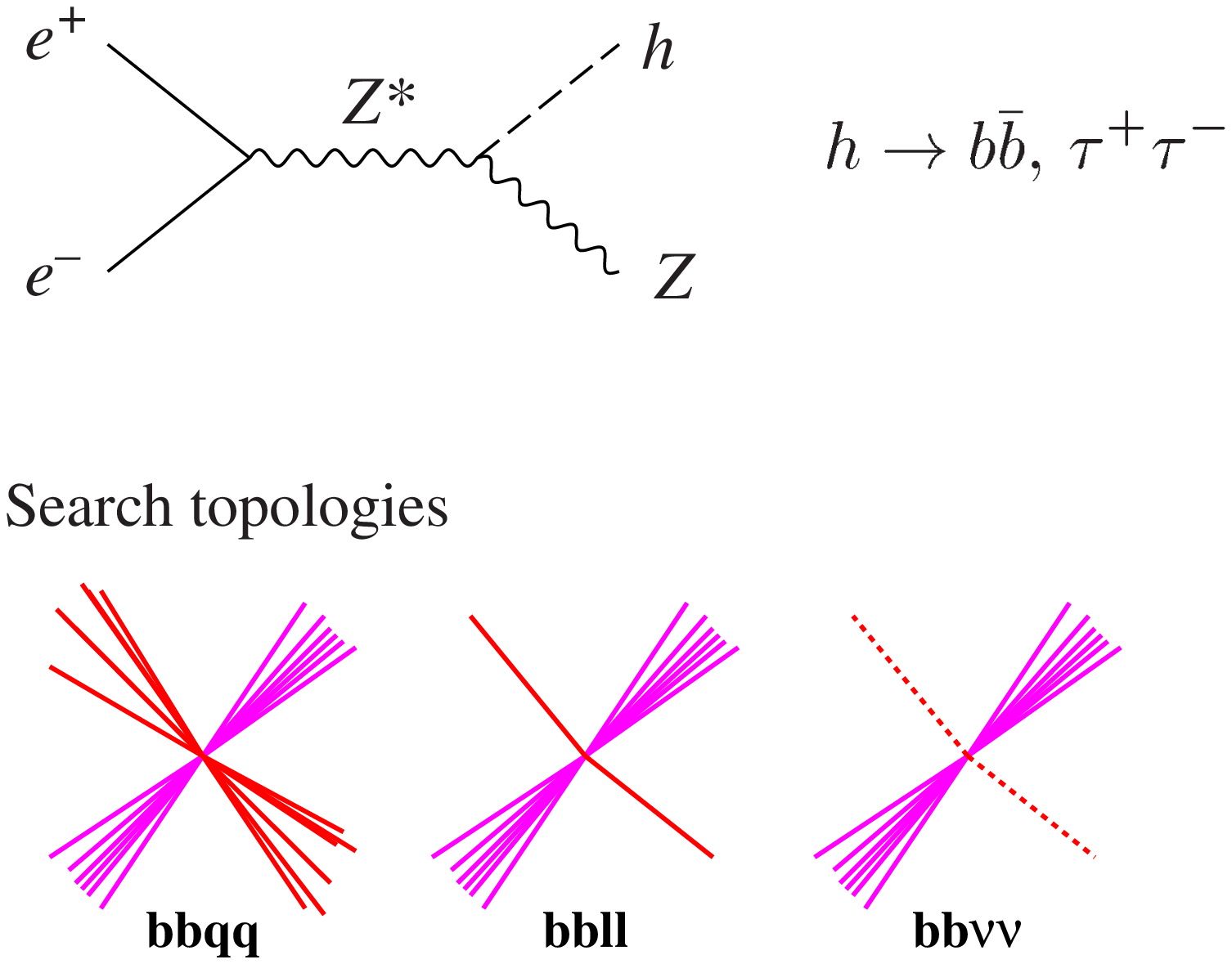}

\caption{Higgs production mechanism at LEP-2 and the final state topologies.}
\end{figure}

\begin{figure}[t]
\centering\leavevmode
\epsfxsize=3in\epsffile{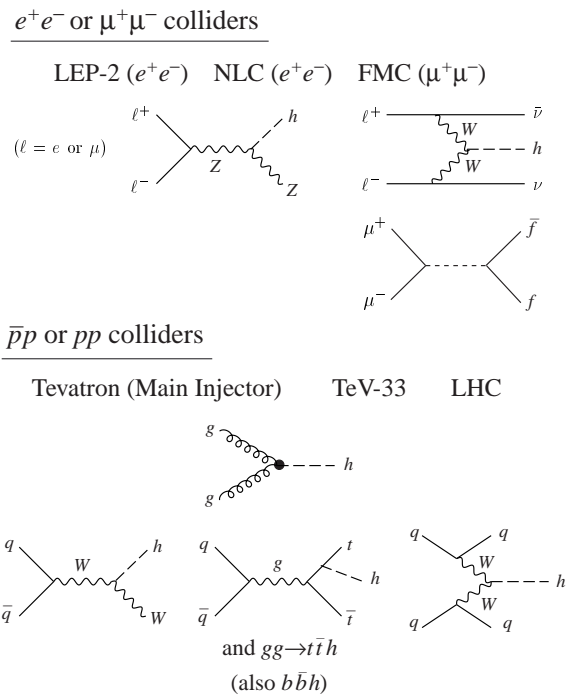}

\caption{Diagrams for Higgs boson production at lepton and hadron colliders.}
\end{figure}

\begin{figure}[t]
\centering\leavevmode
\epsfxsize=3in\epsffile{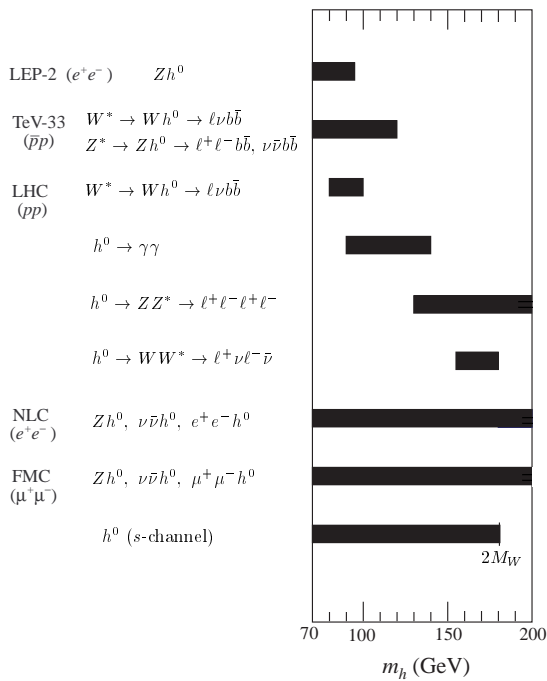}

\caption{SM Higgs intermediate mass ranges for which signals can be detected at future colliders.}
\end{figure}

\begin{figure}[t]
\centering\leavevmode
\epsfxsize=2.5in\epsffile{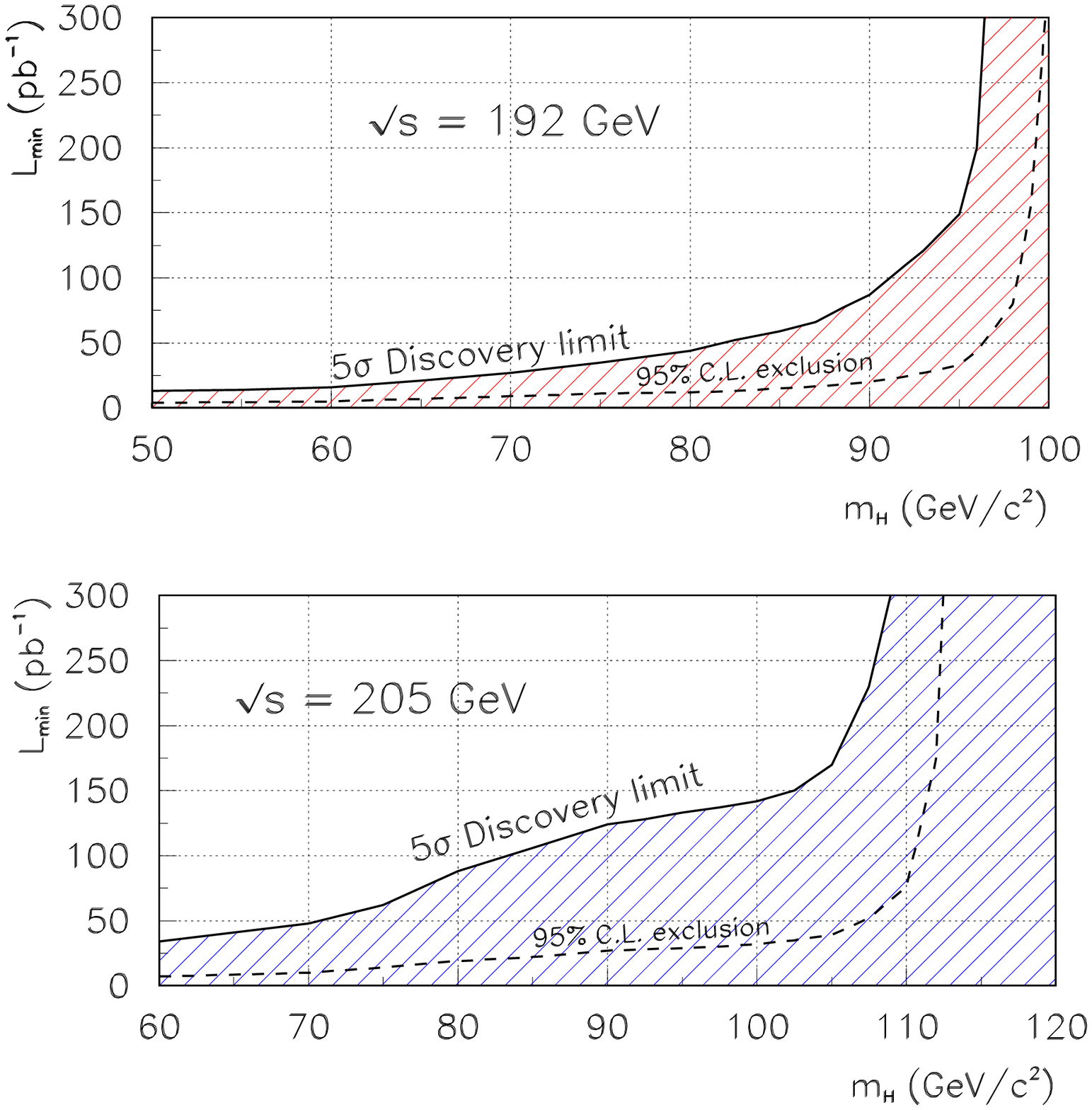}

\caption[]{Minimum luminosity per LEP-2 experiment for $5\sigma$ discovery or exclusion. From Ref.~\citenum{bighiggs}.}
\end{figure}

\section{The Supersymmetry Extension}

There is a serious problem with the SM in that loop corrections to the Higgs mass diverge quadratically, with a cut-off $\Lambda \approx 10^{19}$~GeV from gravity. Two solutions are advanced to remedy this disease. One is to replace elementary Higgs with a bound state, but at present there is no fully satisfactory model of this type. The other is to have a supersymmetry (SUSY) with new particles differing in spin by $|\Delta j| = 1/2$ from SM particles, cancelling the divergent loop contributions to $m_h$. This solution is natural if the mass scale for SUSY particles is $\alt 1$~TeV.

The minimal supersymmetric standard model (MSSM) has double the number of the presently known particles and two Higgs doublets. It has ``soft'' SUSY breaking masses that originate from a hidden sector of the theory; these soft mass parameters consist of all gauge-invariant SUSY breaking terms that do not cause quadratic divergences.\cite{reviews} An approximate decoupling of SUSY radiative contributions to precision electroweak observables maintains the success of the SM, as illustrated\cite{cdf} in Fig.~5.

\begin{figure}[t]
\centering\leavevmode
\epsfxsize=2.5in\epsffile{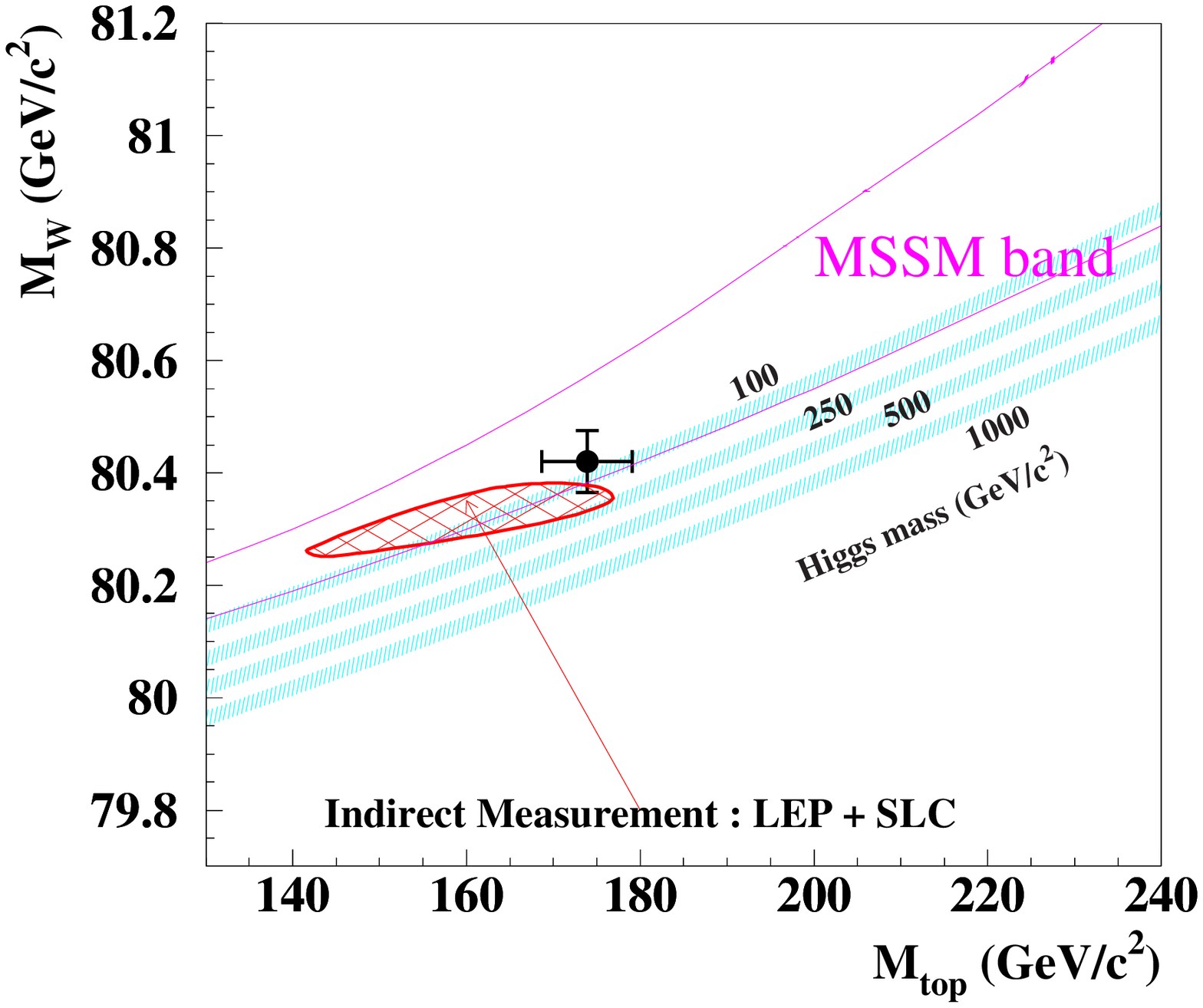}

\caption{MSSM radiative corrections band in $M_W$ versus $m_t$. From Ref.~\protect\citenum{cdf}.}
\end{figure}

The unification of gauge couplings at the scale of a Grand Unified Theory (GUT) is realized for the MSSM (but not the SM). The vacuum polarizations cause the $g_1$, $g_2$, and $g_3$ couplings to evolve differently with the energy scale $\mu$,
\begin{equation}
{dg_i\over d\ln\mu} = {b_i g_i^3\over 16\pi^2} \,,
\end{equation}
and an approximate intersection of couplings occurs at $\mu\simeq 10^{16}$~GeV, as illustrated in Fig.~6 for an effective SUSY scale of 1~TeV. Thus the SM groups may be subgroups of a GUT group such as SU(5), SO(10), or E$_6$. The $N=1$ supergravity (SUGRA) model unifies the electroweak, strong, and gravitational interactions. SUSY breaking is communicated from the hidden sector via gravitational interactions, giving universal soft parameters at the GUT scale (with small Planck scale induced corrections possible\cite{non-u}). Starting with a universal fermion mass ($m_{1/2}$) and a universal scalar mass ($m_0$) at the GUT scale, the masses and couplings can be evolved to the electroweak scale using the Renormalization Group Equations (RGEs).

\begin{figure}
\centering\leavevmode
\epsfxsize=3.25in\epsffile{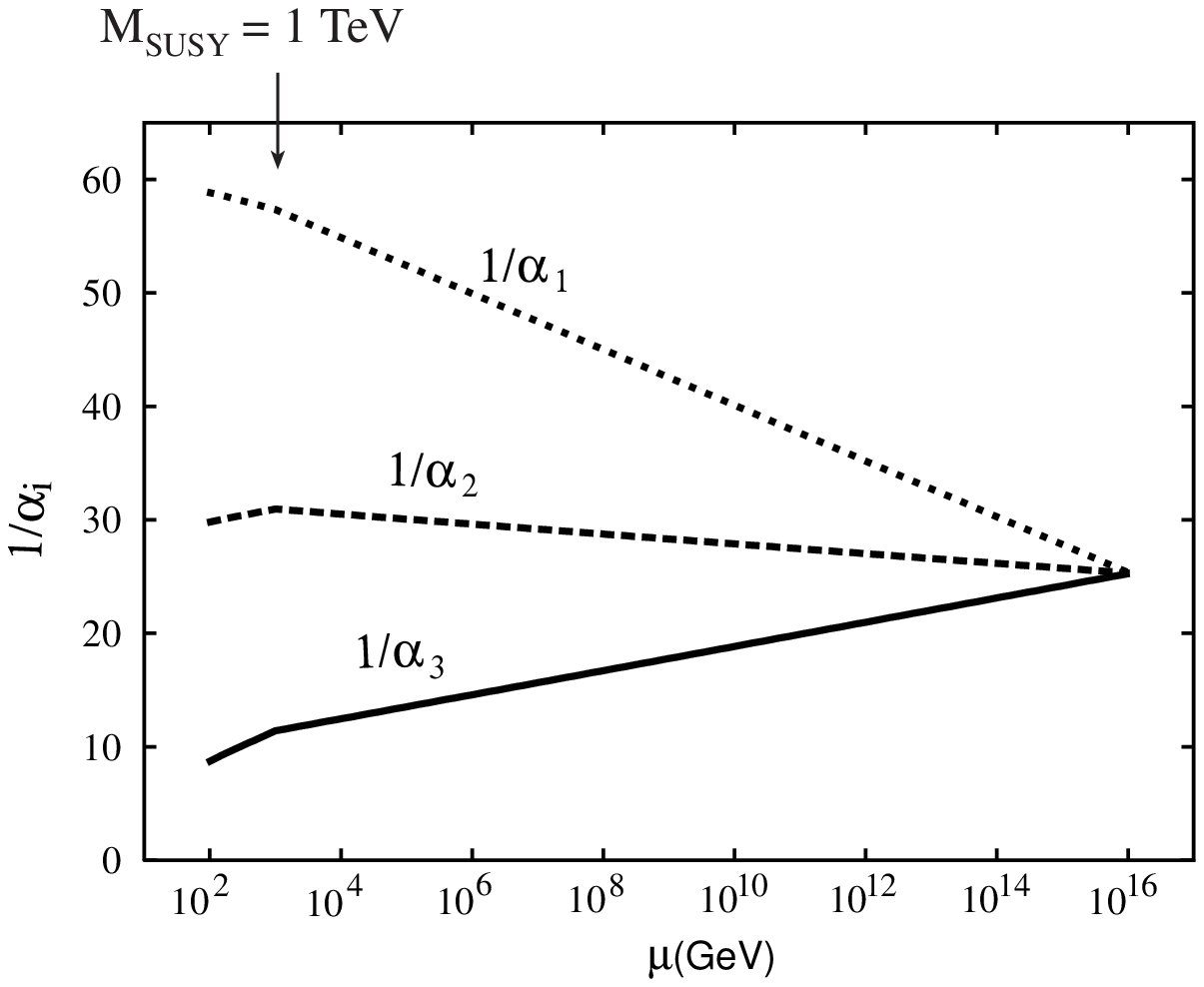}

\caption{Unification of gauge couplings in the MSSM with an effective SUSY scale of TeV.}
\end{figure}

Mass generation in the MSSM occurs via two Higgs doublets $(H_u^+, H_u^0)$ and $(H_d^0, H_d^-)$ through Yukawa couplings to the fermions. The Higgs miracle is explained by having a large $H_u^0 \bar tt$ coupling $\lambda_t$ at the GUT scale that drives $M_{H_u}^2 < 0$ at the electroweak scale. The mass ratio $m_b/m_\tau$ is also correctly predicted in the MSSM (but not in the SM) with a large $\lambda_t$ at $M_{\rm GUT}$.

In its evolution to low energies, $\lambda_t$ approaches a quasi-fixed point, independent of its precise value at the GUT scale so long as $\lambda_t$ is large at $M_{\rm GUT}$. The top quark mass is given by $m_t = \lambda_t \sin\beta {v\over\sqrt2}$, where $\tan\beta = v_u/v_d$, and the fixed point value\cite{bbo,carena} is $\lambda_t = 1.06$ for low $\tan\beta$. Thus the prediction is $m_t = (200\rm~GeV)\sin\beta$ and the measured $m_t = 175$~GeV implies $\tan\beta \simeq 1.8$. More generally there is another solution with large $\tan\beta \simeq 56$; see Fig.~7.

\begin{figure}
\centering\leavevmode
\epsfxsize=3.25in\epsffile{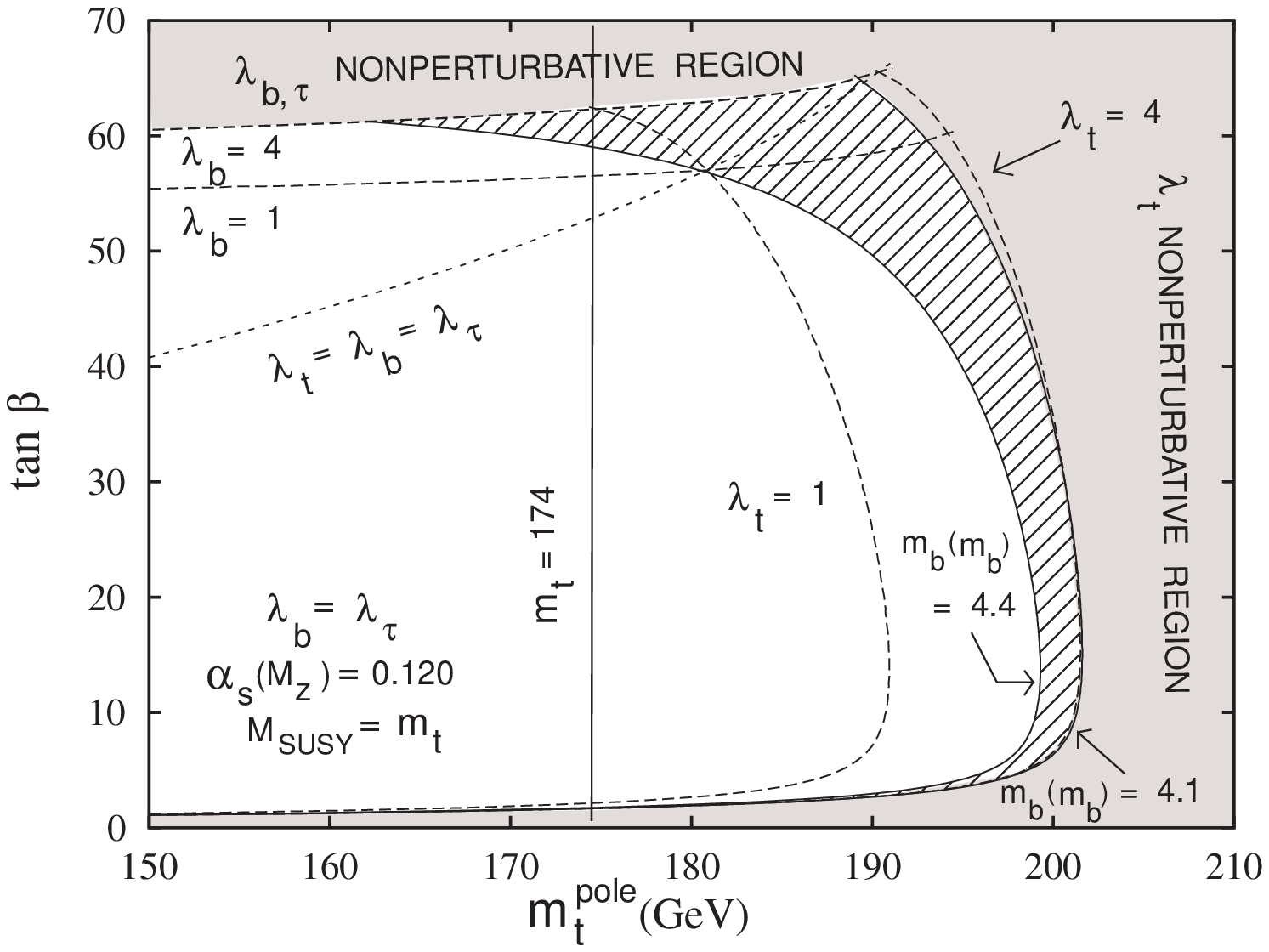}

\caption{Contours of constant $m_b(m_b)$ in the $m_t(\rm pole)$ versus $\tan\beta$ plane; contours of constant GUT scale couplings are also shown. From Ref.~\protect\citenum{bbo}.}
\end{figure}

\section{Light Higgs in Supersymmetry}

The MSSM Higgs sector has 5 physical Higgs bosons: $h^0,\ H^0,\ A^0,\ H^\pm$. Two parameters, $\tan\beta$ and $m_A$, specify all the Higgs masses and couplings at tree level. Because the Higgs self-coupling is related to the gauge coupling in supersymmetry, there is a tight upper limit on the mass of the lightest Higgs boson. For $m_A\gg M_Z$, the limit is\cite{carena2}
\begin{equation}
m_h^2 \simeq M_Z^2 \cos^2 2\beta + {3G_F\,m_t^4\over \sqrt 2 \pi^2} \log{m_{\tilde t}^2\over m_t^2} \,,
\end{equation}
where $\tilde t$ denotes the scalar partner of the top quark. The upper limit on $m_h$ is reached for $m_A \agt 250$~GeV. For large $\tan\beta$ the upper bound is $m_h \alt 130$~GeV. In GUTs with $b$-$\tau$ unification the low $\tan\beta$ infrared fixed point gives the bound $m_h \alt 100$~GeV. The lightest Higgs boson is the jewel in the SUSY crown; it is a secure experimental target of a low-energy SUSY. The low values of $m_h$ inferred from precision electroweak data may well be the ``smoking gun'' for supersymmetry. The current direct search limits from LEP-2 give $m_h \agt 75$~GeV. The combination of $h^0Z$ and $h^0A^0$ searches at LEP-2 will test the $\tan\beta = 1.8$ infrared fixed point solution in the near future.\cite{falk} 

\section{Muon Colliders for Higgs Physics}

Once the Higgs discovery is made, at LEP-2, the Tevatron, or the LHC, how can we best study its properties? At a muon collider, the Higgs boson can be produced as an $s$-channel resonance,\cite{bbgh-prl} with a cross section that is $(m_\mu/m_e)^2 \approx 40{,}000$ times that for $e^+e^-$ collisions, offering unique opportunities. Muon colliders are believed to be feasible.\cite{mumu-procs} Protons on a target yield pions whose decay muons can be cooled via ionization media, then rapidly accelerated and stored in a circular ring where ${\approx}10^3$ crossings take place before decay. The advantages over $e^+e^-$ colliders are suppressed synchrotron radiation ($\propto 1/m^4$) and smaller overall size. Muon colliders would provide sharp beam energy ($\sigma_{E_{\rm c.m.}} \sim 2$~MeV) with small initial state radiation, no beamstrahlung, and precise energy calibration (to 1~ppm) through spin rotation measurements. These properties are essential for precision Higgs mass and width studies. With prior study of $h$ at the LHC, we will know $m_h$ to an accuracy $\Delta m_h \sim 100$~MeV. Then the muon collider ring can be designed with $\sqrt s = m_h$.

As an example, consider a Higgs mass $m_h = 110$~GeV. At $\sqrt s = 110$~GeV the $b\bar b$ signal ${}\simeq 10^4$ events/fb and the $b\bar b$ background ${}\approx 10^4$ events/fb. There is a strong $\sqrt s$ dependence of the background due to the proximity of the $Z$-resonance pole. With an integrated luminosity $L \approx 1.5\times 10^{31}\rm\,cm^{-2}\,s^{-1}$ ($0.15\rm~fb^{-1}$), a scan for $h^0$ determines $\Delta m_h \sim 1$~MeV in one year; the results of a simulated scan are shown in Fig.~8. With a subsequent fine scan over the $h$-resonance peak, the following accuracies could be achieved with 0.4~fb$^{-1}$ integrated luminosity:
\begin{equation}
\Gamma_h^{\rm tot}{:}\ 16\% \qquad \sigma{\rm BF}(b\bar b){:}\ 1\% 
\qquad \sigma{\rm BF}(WW^*){:}\ 5\%
\end{equation}
From these measurements the mass $m_A$ can be inferred,\cite{bbgh-pr} provided that $m_A \alt 450$~GeV.

\begin{figure}
\centering\leavevmode
\epsfxsize=3in\epsffile{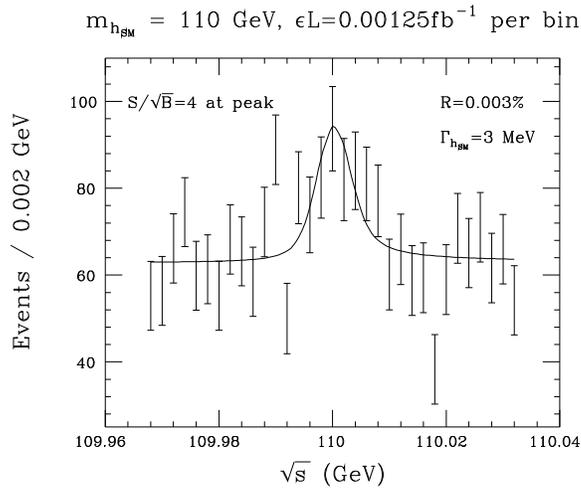}

\caption[]{Number of $b\bar b$ events versus $\sqrt s$ in the vicinity of $m_h=100$~GeV, for a muon beam resolution of 0.03\%. From Ref.~\protect\citenum{bbgh-pr}.}
\end{figure}

\section{Supersymmetric Particle Searches}

The table below lists the SM particles and their MSSM companions.
\begin{table}[t]
\caption{Standard Model particles and the corresponding supersymmetry particles.}
\medskip
\centering\leavevmode
\epsfxsize=1.9in\epsffile{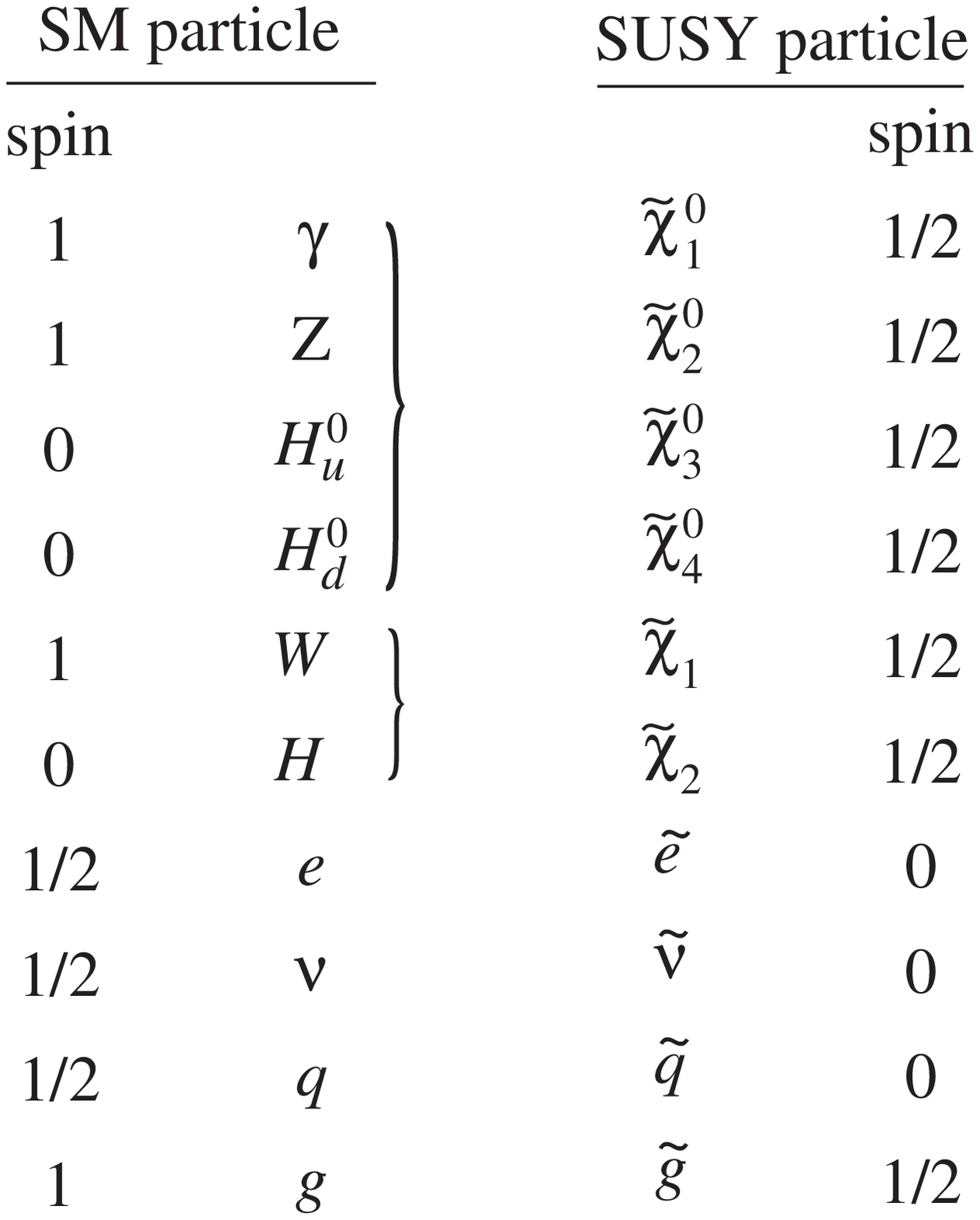}
\end{table}
The colored SUSY particles ($\tilde q, \tilde g$) are expected to be heavier than the color singlet particles. The states $\tilde\chi_1^0, \tilde\chi_2^0, \tilde\chi_1^\pm, \tilde\ell, h$ are typically ``light", while $\tilde\chi_3^0, \tilde\chi_4^0, \tilde\chi_2^\pm, \tilde g, \tilde q$ are ``heavy''. Figure~9 shows the mass spectra of light states versus the universal scalar mass $m_0$ at the GUT scale in the SUGRA model. The masses approximately satisfy the scaling ratios\cite{arno2} $\tilde\chi_1^0 : \tilde\chi_2^0 : \tilde\chi_1^\pm : \tilde g = 1:2:2:7$. At LEP-2 a lower bound $m(\tilde\chi_1^\pm) \agt 91$~GeV has been placed on the lightest chargino mass. On the upper horizontal axis of Fig.~9, the values of the relic density for neutralino dark matter are given.\cite{bk} The acceptable range $0.1\alt \Omega h^2 \alt 0.4$ is indicated by the shaded region.

\begin{figure}[t]
\centering\leavevmode
\epsfxsize=3.2in\epsffile{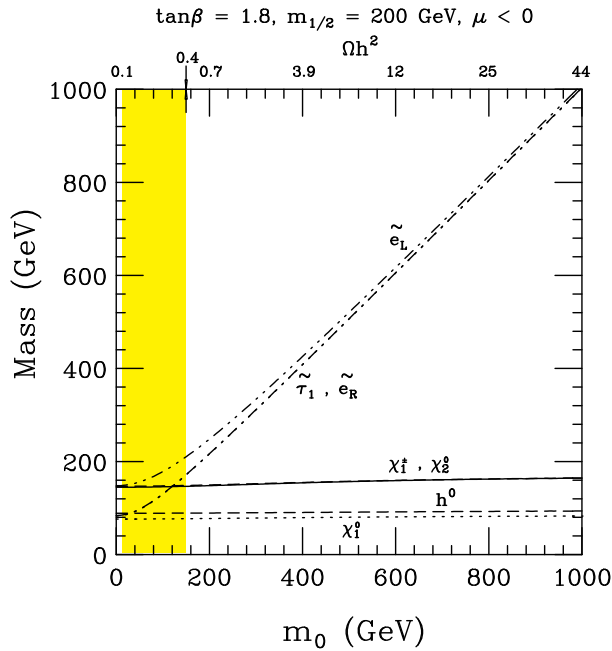}

\caption{Mass spectra of light SUSY particles in the supergravity model versus the universal scalar mass $m_0$.}
\end{figure}

A new conserved number, $R$-parity, is introduced in supersymmetry to keep the proton sufficiently stable; $R=+1$ for normal particles and $R=-1$ for SUSY particles. The lightest SUSY particle (LSP) is then stable. In SUGRA models the neutralino $\tilde\chi_1^0$ is the LSP. Other SUSY particles decay to $\tilde\chi_1^0$, which is undetected and thus a source of missing energy in collider events.

A classic SUGRA signature at the Tevatron or the LHC are trileptons due to the subprocess\cite{arno3,baer,bkl}
\begin{equation}
q\bar q \to \tilde\chi_1^+ \tilde\chi_2^0 \,,
\end{equation}
with the decays $\tilde\chi_1^+\to\tilde\chi_1^0 \ell^+ \nu$ and $\tilde\chi_2^0\to\tilde\chi_1^0\ell^+\ell^-$. Figure~10 shows the recent calculation\cite{bkl} of the cross section for trileptons at the Main Injector upgrade of the Tevatron ($\sqrt s = 2.0$~TeV). A luminosity of 2 to 4~fb$^{-1}$ will be accumulated by both CDF and D0\llap/ detectors. Interesting regions of SUGRA parameter space will be explored.

\begin{figure}[t]
\centering\leavevmode
\epsfxsize=3in
\setbox0\hbox{\epsffile{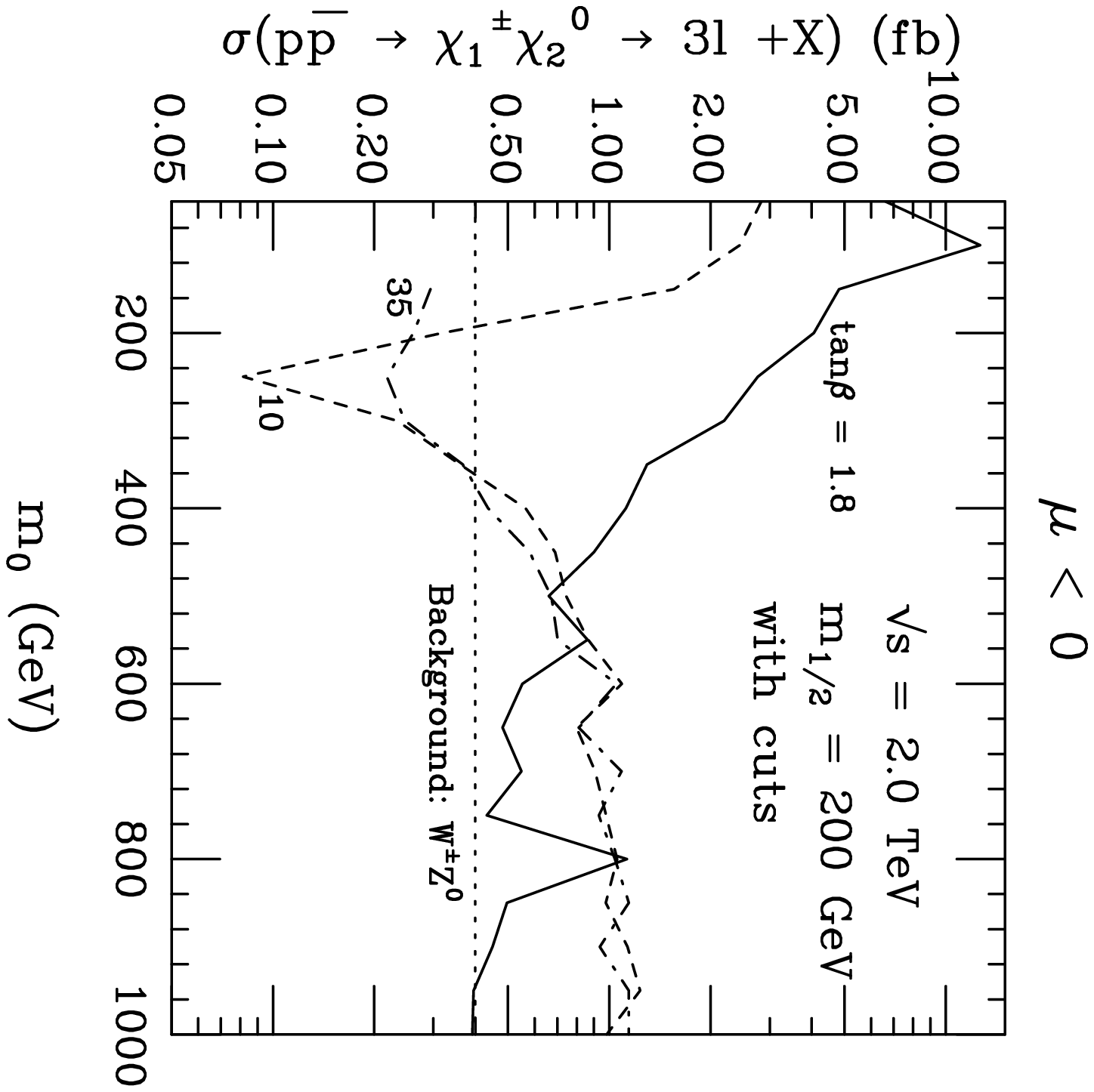}}
\rotl0

\caption[]{Cross section for production of trilepton events via $\chi_1^\pm\chi_2^0$ at the Tevatron for $\sqrt s = 2.0$~TeV. From Ref.~\citenum{bkl}.}
\end{figure}

The other classic SUGRA signatures are missing transverse energy ($\E) +\rm jets$ and same-sign ${\rm dileptons} + \E + \rm jets$ due to strong production of $\tilde g\tilde g$, $\tilde g\tilde q$, and $\tilde q\tilde q$. The $\tilde\chi_1^0$ emitted in decays is the source of $\E$. The $\tilde g$ is a Majorana particle so decays to $\tilde\chi^+$ and $\tilde\chi^-$ have equal rates and lead to same-sign dileptons.\cite{bkp} From searches for the above SUSY signals by the CDF and D0\llap/ collaborations exclude $m_{\tilde g} = m_{\tilde q} \alt 250$~GeV.  The LHC has discovery reach sufficient to find SUSY particles within the natural scale $M_{\rm SUSY} \alt 1$--2~TeV.

\section{Summary}

The mysteries of electroweak symmetry breaking and mass generation are about to be solved by forthcoming collider experiments. Precision electroweak tests point a smoking gun towards the lightest supersymmetric Higgs boson $h^0$. It is still a horse race to the discovery line for $h^0$, with the contenders being LEP-2 (where energy is critical and the infrared fixed point prediction will be tested soon), the Tevatron (where luminosity is critical), and the LHC (where no difficulty is expected in finding $h^0$). The LHC and possibly the Tevatron can do ``the job'' of finding the particles of supersymmetry and determining their properties. Lepton colliders ($\mu^+\mu^-$ and $e^+e^-$) are needed for precision studies of supersymmetric particles. 

Supergravity models are the paradigm for supersymmetry breaking, but gauge-mediated messenger models are also receiving great attention. The experimental signatures of gauge-mediated models are different from those of SUGRA. Eventually M-theory may give insights on the nature of supersymmetry breaking and of the physics at the GUT and Planck scales.

\section*{Acknowledgements}
I thank Chung Kao for helpful comments in the preparation of this report.
This work was supported in part by the U.S.~Department of Energy under Grant No.~DE-FG02-95ER40896 and in part by the University of Wisconsin Research Committee with funds granted by the Wisconsin Alumni Research Foundation.

\end{document}